\documentclass[graybox]{svmult}

\usepackage{mathptmx}       
\usepackage{helvet}         
\usepackage{courier}        
\usepackage{type1cm}        
\usepackage{makeidx}         
\usepackage{graphicx}        
\usepackage{multicol}        
\usepackage[bottom]{footmisc}

\makeindex             

\begin{document}

\title*{The spectroscopic observations of CoRoT asteroseismic targets with HARPS}
\titlerunning{The spectroscopic observations of CoRoT asteroseismic targets with HARPS}
\author{Poretti, E., Rainer, M., Mantegazza, L., Floquet, M.,
Mathias, Ph., Amado, P., 
Aerts, C., Uytterhoeven, K.  and the CoRoT Seismology Ground-Based Working Group}
\authorrunning{Poretti et al.} 
\institute{
Poretti Ennio,  Rainer Monica, Mantegazza Luciano \at INAF-OA Brera, Via E. Bianchi 46, 23879 Merate, Italy. 
\and Floquet Michele \at GEPI, Observatoire de Paris,  5 place Jules Janssen, 92195 Meudon Cedex, France
\and Mathias Philippe \at IRAP, 
57 Avenue d'Azereix, 65008 Tarbes, France
\and Amado Pedro \at Instituto de Astrof\'{\i}sica de Andaluc\'{\i}a, Apartado 3004, 18080 Granada, Spain
\and Aerts Conny \at Instituut voor Sterrenkunde, KU Leuven, Celestijnenlaan 200D, 3001, Leuven, Belgium 
\and Uytterhoeven Katrien \at 
Instituto de Astrof\'{\i}sica de Canarias (IAC), Tenerife, Spain
}
%
%
\maketitle

\abstract*{
CoRoT photometric measurements of asteroseismic targets need complementary ground-based  
spectroscopic observations.  
We are using the planet-hunter HARPS  spectrograph attached to the  3.6m-ESO telescope 
in the framework of two consecutive Large Programmes. 
We discuss its use to study line-profile variations and we report on a specific 
result obtained for the $\delta$ Sct star HD~170699.
}
\abstract{
CoRoT photometric measurements of asteroseismic targets need complementary ground-based  
spectroscopic observations.  
We are using the planet-hunter HARPS  spectrograph attached to the  3.6m-ESO telescope 
in the framework of two consecutive Large Programmes. 
We discuss its use to study line-profile variations and we report on a specific 
result obtained for the $\delta$ Sct star HD~170699.
}

\section{Introduction}

Ground-based spectroscopic observations
began at once  with the launch of the CoRoT satellite.
They involve several high-resolution \'echelle spectrographs: FEROS and 
HARPS at ESO-La Silla (Chile), FOCES at Calar Alto (Spain), SOPHIE at the 
Observatoire Haute Provence (France), FIES at NOT and HERMES at Mercator \,\cite{hermes} 
(both at 
the Observatorio Roque de los Muchachos, La Palma, Spain). A small number 
of spectra were also obtained with CORALIE at ESO-La Silla and 
HERCULES at Mt. John Observatory (New Zealand).
After the 
completion of the ESO Large Programme with FEROS (LP 178-D.0361, 60 nights in total), 
we obtained two subsequent Large Programmes with HARPS, the famous
planet--hunter \'echelle spectrograph
primarily used for the measurements of radial velocities. 
The first Large Programme with HARPS (LP 182.D-0356) consisted  of 
45 nights of observations between December 2008 and December 2009, 
while the  current LP 185.D-0056 started on June 2010 and will end 
in January 2013. As of now, we have taken about 2500 HARPS spectra 
of CoRoT asteroseimic targets.
 The high-resolution spectra make the
detection of high-degree modes possible \cite{bvcir} since the stellar disk can be spatially
resolved thanks to the Doppler shifts induced by the star's rotation. Therefore,
by measuring the observed variations in the line profiles, it is possible to
 know what kind of modes are
excited in the stars and to assign the spherical wavenumbers $(\ell, m)$ to each
of them. 
The HARPS spectra are reduced and normalized in a homogeneous way using a
semi-automated pipeline developed at INAF-OAB. 

\section{HARPS: improvements in the instrument setup}

HARPS is able to operate in two configurations: 
high accuracy (HAM mode, with R=115,000) or high efficiency 
(EGGS mode, with R=80,000). 
At the telescope, the HARPS pipeline provides an estimate of the signal--to--noise
ratio (SNR) at three different wavelengths (4500, 5000, and 6500 \AA). We
 chose to work mostly in the 
EGGS configuration, aiming at a signal-to-noise ratio (SNR) estimated 
at the telescope of about 200. In the reduction process we computed 
the SNRs taking into account photon noise, readout noise, and flat field
correction. We could infer that for our stars (B-A-F spectral types)
the SNR values given by the HARPS pipeline at 5500\,\AA\, are a little too
optimistic when compared with the values we measured between 5802 and
5825\,\AA. The ratio between the two SNR values is around 1.23 for A-F stars
and around 1.28 for B stars.

\begin{figure}[]
\includegraphics[angle=270,scale=0.4]{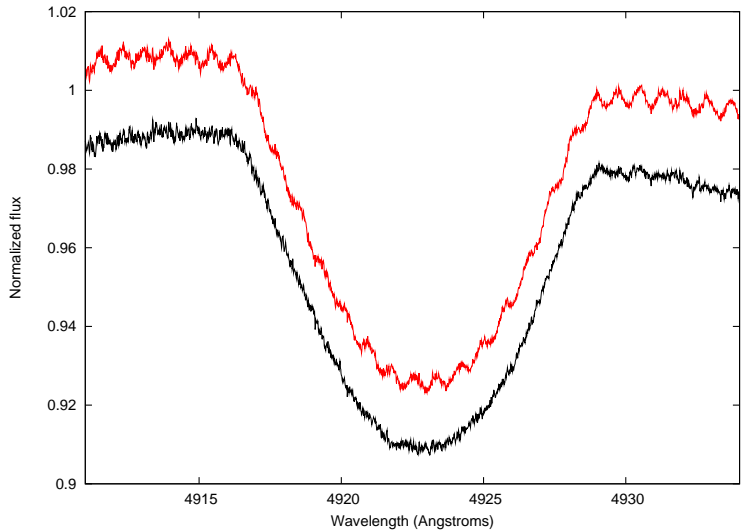}
\caption[t]{Average 4921\,\AA\, He\,I line of the Be star HD 51452: the instrumental effect  
(upper red curve) was eliminated by
removing the blue filter located on the optical path of the HARPS calibration lamp (lower black curve).}
\label{be}
\end{figure}

Analysing the HARPS data obtained during the first observing runs 
(14-24 December 2008 and  3-8 January 2009) of the Large Programme 
ESO LP 182.D-0356, we found the presence of a disturbing  feature in the 
spectra, i.e., an oscillation with an amplitude of about 0.05\% of the continuum.
It was the first time that this problem was reported. In fact, our approach 
to the data is different than the one usually used with the HARPS spectra. 
Usually the spectra are used to compute radial velocities of cool stars, 
while we search for  line profile variations in hot stars. 
In particular, this feature became evident only by averaging a large number 
of spectra of the same line (Fig.~\ref{be}, upper red curve), while it is practically 
invisible in a single spectrum.
During the observing run of June 2009, several attempts were made to 
understand the nature of this problem, which in the meantime had worsened. 
It was found that the filters in front of the calibration lamps were misaligned 
after a check-up in May 2009, and this enhanced the amplitude of the 
oscillations, which reached a value of about 2\% of the continuum.
Realigning the filters decreased the feature's amplitude to the previous 0.05\% 
of the continuum, but did not eliminate it. With the help of the ESO staff 
we were able to identify the source of this problem as an interference created 
by the flat-field filters (red and blue). In fact, the issue was 
solved by taking away the blue filter: the subsequent flat-field correction
was free from the instrumental effect (Fig.~\ref{be}, lower black  curve).

\section{Evidence of high--degree modes excited in $\delta$ Sct stars}
Several papers were published using the spectra observed in the ground-based programme 
(see the special A\&A vol. 506) and 
a lot of work is still in progress. The data have been used in several ways, 
ranging from frequency analysis to the identification of the physical parameters of the stars.
\begin{figure}[]
\includegraphics[scale=0.5]{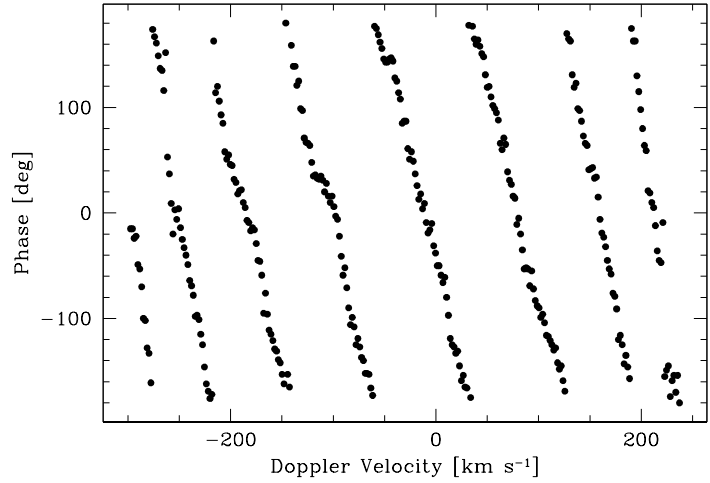}
\caption[t]{HD 170699: phase variations across the line profile of the  $f$=48.26~d$^{-1}$ term
(HARPS spectra)}
\label{phase}
\end{figure}
We present here the case of the $\delta$ Sct stars, which
 are main--sequence or slightly evolved stars showing  multiperiodic
light curves. 
CoRoT introduced a dramatic change in terms of  both  number of identified
frequencies and of the involved physics:  hundreds of frequencies were detected
in the light curve of HD 50844 \cite{hd50844}.
The main goal of the spectroscopic observations is to 
determine if the observed frequencies  can be associated with pulsation
modes. 
The mode identification was performed by fitting the amplitude and phase variations
of each mode across the average line profile of each HARPS spectrum \cite{vienna},
\cite{zima}.  
Figure~2 shows the phase diagram  of the high--frequency
term $f$=48.26~c/d detected in the HARPS spectra of HD 170699. 
The total amount of the phase shifts suggests that it should be  a $\ell$=16$\pm$1 mode
\cite{telt}. This frequency is the second detected in the spectroscopic
data and  the 374$^{\rm th}$ in the CoRoT photometric light curve, with an amplitude
of about 50\,$\mu$mag. Therefore, the case of HD 170699 seems to support the
conclusion made in the case of HD 50844 \cite{hd50844}, i.e., that cancellation
effects are not sufficient in completely removing the variations of the integrated flux
associated
with $\ell\ge$5  modes. Two other  physical mechanisms  can be invoked to explain
the rich frequency spectra observed in $\delta$ Sct stars: granulation effects 
\cite{kall} or island modes, e.g., \cite{island}. These three different scenarios 
contribute to make the modeling of $\delta$ Sct stars  a challenging task for theoreticians. 
In particular, the visibility  of island modes may be particularly relevant in the case of HD 170699.
We measured $v_{\rm eq}\sin i=270$~km\,s$^{-1}$, a value very close to the break-up velocity, and  
we can infer that HD 170699 is seen equator-on.

\end{document}